\documentstyle[12pt,epsfig]{article}
\def\lsim{\mathrel{\rlap{\lower3pt\hbox{\hskip0pt$\sim$}}
    \raise1pt\hbox{$<$}}}         
\def\gsim{\mathrel{\rlap{\lower4pt\hbox{\hskip1pt$\sim$}}
    \raise1pt\hbox{$>$}}}         

\def\beq{\begin{equation}}   \def\eeq{\end{equation}}
\def\be{\begin{equation}}   \def\ee{\end{equation}}
\def\bea{\begin{eqnarray}}   \def\eea{\end{eqnarray}}

\newcommand{\bibit}[1]{\bibitem{#1}}

\newcommand{\La}{{\,\overline{\!\Lambda\!}\,}}
\newcommand{\Lam}{\Lambda_{\rm QCD}}
\newcommand{\vep}{\varepsilon}

\newcommand{\matel}[3]{\langle #1|#2|#3\rangle}

\newcommand{\aver}[1]{\langle #1\rangle}

\newcommand{\GeV}{\,\mbox{GeV}}
\newcommand{\MeV}{\,\mbox{MeV}}

\begin{document}

\thispagestyle{empty}

\begin{flushright}
UND-HEP-02-BIG\hspace*{.15em}06\\
Bicocca-FT-02-2\\
hep-ph/0202175\\
\end{flushright}
\vspace{.3cm}
\begin{center} \Large
{\bf ~{\hspace*{-4.5mm}On The Expected Photon Spectrum in
\boldmath $B \!\to\! X_s \!+\! \mbox{{\LARGE$\gamma$}} $
\hspace*{-4.5mm}~\\and Its Uses}}\\
\end{center}
\vspace*{.3cm}
\begin{center} {\Large Ikaros Bigi $^{a\,\alpha}$\\
Nikolai Uraltsev $^{a,b,c}$}\\
\vspace{.4cm}

{\normalsize
{\it ~\hspace*{-17mm}{\rm $^a$}Department of Physics,
University of Notre Dame du Lac,
Notre Dame, IN 46556 USA\hspace*{-15mm}~}
}\\
$^b${\it INFN, Sezione di Milano, Milan, Italy\\
$^c${\it Petersburg Nuclear Physics Institute, Gatchina,
St.\,Petersburg, 188350, Russia}
}
\vspace{.3cm}
{\small $^\alpha$\,{\tt e-mail address:} {\sf bigi.1@nd.edu\\
}}
\vspace*{15mm}
{\small {\bf Contributed to the Workshop on the CKM Unitarity Triangle\\
CERN, February 13-16, 2002}}
\vspace*{15mm}

{\Large{\bf Abstract}}\\ 
\end{center}
Measuring the photon energy spectrum in radiative $B$ decays provides 
essential help for gaining theoretical control over semileptonic $B$ 
transitions. The hadronic recoil mass distribution in  $B \!\to\! X_u
\,\ell \nu $ promises the best environment for  determining 
$|V_{ub}|$. The theoretical uncertainties are largest in the domain
of low values of the lepton pair mass $q^2$.  Universality relations
allow to describe this domain reliably in  terms of the photon spectrum
in $B \!\to\! X_s \!+\! \gamma$. A method is proposed to incorporate $1/m_b$
corrections into  this relation. The low-$E_\gamma$ tail in radiative 
decays is important in the context of extracting $|V_{ub}|$. 
We argue that CLEO's recent fit to the spectrum underestimates
the fraction  of the photon spectrum below $2\GeV$. Potentially
significant uncertainties enter in the theoretical evaluation of the
integrated end-point lepton spectrum or the $B \!\to\! X_u \,\ell \nu $
width with a too high value of the lower cut on $q^2$ in alternative
approaches to $|V_{ub}|$.  
\vfill

\setcounter{page}{0}

\newpage
\tableofcontents
\vspace*{2mm}

\section{Introduction}

Since $B \to X \!+\! \gamma$ decays represent a loop
effect within the Standard Model (SM), they are viewed as a window
onto New Physics. There is a second motivation that has attracted
considerable attention: measuring the inclusive photon spectrum
allows us to infer the motion of the heavy quark inside the $B$
meson. This yields new insights into the inner
workings of QCD which significantly affect, for example, the lepton
spectrum in $b\to u \,\ell\nu$ decays.
The moments of the heavy quark distribution function allow in principle
to determine experimentally a number of nonperturbative 
parameters intrinsic to the
heavy quark expansion the knowledge of which is of primary practical
importance \cite{prl,optical}.

One should note, however, that we are not in the dark about the values of 
these
quantities: we know the running beauty quark mass $m_b$ and the kinetic
$\mu_{\pi}^2$ and chromomagnetic $\mu _G^2$ expectation values with good 
accuracy due
to intensive theoretical investigations over the last few years (for a review 
see
Ref.~\cite{ioffe}). The most accurate value for $m_b$ has been determined from 
$e^+e^-$
annihilation to beauty hadrons close to
threshold. Heavy quark sum rules impose tight constraints; in particular they 
ensure
$\mu _\pi ^2 \!>\! \mu _G^2$, which in practice allows for only a
limited range for the former. Extracting the values of $m_b$, $\mu _\pi ^2$ 
etc. from
the moments or shape of energy spectra thus serves as a cross
check -- and a highly valued one at that -- of our theoretical control. In 
lucky
cases this can allow to narrow down the
theoretical uncertainties by excluding some of the corners in parameter space.
This comparison provides us also with information about the scale of higher 
order
effects which will be very important in assessing the numerical
accuracy of concrete applications of the heavy quark
expansion \cite{CLEOhadr}.

It has always been tempting to extract $|V_{ub}|$ from the lepton energy 
endpoint
spectrum in $B \to X_u\, \ell \nu $ using information on the $b$ quark
distribution function inferred from the photon spectrum in 
$B \to X_s\! +\!\gamma$. We will
illustrate that theoretical problems arise in such a procedure which may 
severely limit its
accuracy. It is more profitable to use the information from the photon 
spectrum to control
possible theoretical uncertainties in the hadronic recoil mass spectrum in $B 
\to X_u\, \ell
\nu $ originating from low-$q^2$ kinematics.

Very recently the CLEO collaboration has presented a new measurement of the 
branching
ratio and photon spectrum for $B \!\to\! X_s\!+\!\gamma$
\cite{CLEO1}. The bulk of their spectrum shown in Fig.~1 is quite similar to 
the
theoretical predictions (Fig.~3 of Ref.~\cite{bsg}) made already in
1995. Since then our knowledge of $m_b$ and $\mu _\pi ^2$ has
become more precise; the value of the latter, which determines
the width of the energy distribution is now expected in the lower half
of the interval used in 1995. This brings the main part of the spectrum into 
even better agreement with the data. One
should keep in mind, though, that the experimental photon spectrum is
additionally Doppler-smeared by the nonvanishing velocity
$\beta \!\approx\! 0.063$ of $B$ mesons in decays of $\Upsilon(4S)$.

\begin{figure}[hhh]
\begin{center}
\vspace*{-3mm}
\mbox{\psfig{file=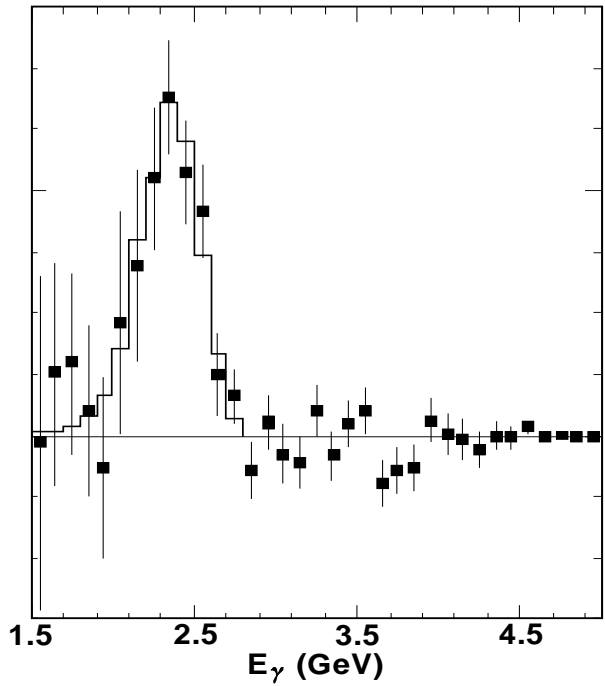,height=9cm,width=8cm}}
\end{center}
\vspace*{-7mm}
\caption{{\small 
CLEO's photon spectrum and the model fit to the data 
from Ref.~\cite{CLEO1}.
} 
}
\end{figure}

The model spectrum used by CLEO (hereafter referred to as CLEO's fit) 
assigns only a tiny fraction of the photon spectrum 
to fall below $2\GeV$. This would be a welcome feature since the 
background becomes much more severe there; it means that the branching 
ratio for radiative decays can therefore be
measured with good accuracy. 
The total radiative branching ratio is little affected by 
the possible uncertainties in this fraction. Yet the
photon spectrum below 2 GeV has considerable intrinsic interest. Since it is 
under better theoretical 
control as far as perturbative and nonperturbative effects are concerned, 
it provides a testbed
for studying the onset of local parton-hadron duality. Its possible
violation, although of no practical concern for integrated widths, can
significantly affect the decay rates integrated over limited kinematic
domains, or interfere with extracting parameters describing
strong dynamics from higher moments. Besides, data from this kinematical 
domain are
important for establishing precision control over the low-$q^2$ region in
$B \to X_u \,\ell \nu  $, referred to above.

It is therefore important to have accurate data on the photon spectrum below 
$2\GeV$ despite the experimental challenge it poses.
CLEO has performed a preliminary measurement of that domain. 
However, we think that the experimental fit in
Ref.~\cite{CLEO1} significantly underestimates the spectrum that has to exist 
there.
The theoretical framework itself used to
translate the observed spectrum into heavy quark parameters assumes
that a more significant fraction of the photon spectrum falls below
$2\GeV$. 
We make the case for future data with even higher statistics to
clarify this issue, and suggest they will show that more than 
$10\%$ of the photon spectrum resides below $2\GeV$. An intriguing lesson can,
however, be drawn if  CLEO's fit were to be confirmed.

The remainder of the paper is organized as follows: after briefly recalling in
Sect.~\ref{EXPECT} how the photon spectrum in radiative $B$ decays is
obtained theoretically,
we discuss more specifically the perturbative spectrum mainly responsible
for the lower tail in Sect.~\ref{PERT}. The complete spectrum is
considered in Sect.~\ref{COMB}. In Sect.~\ref{SKEPT} arguments are given 
calling for caution
when connecting the lepton energy endpoint spectrum in 
$B \to X_u\, \ell \nu $ with the photon
spectrum in $B \to X_s\!+\!\gamma$, and a simplified method to account 
for a part of subleading
corrections is suggested.
We then describe a way to further reduce the theoretical uncertainties in 
extracting
$|V_{ub}|$ from the hadronic recoil mass spectrum in 
$B \to X_u\, \ell \nu $ in Sect.~\ref{NEW},
before commenting on some issues related to extracting $|V_{ub}|$ and 
presenting our conclusions in Sect.~\ref{CONCL}.

\section{Theoretical expectation for the photon spectrum}
\label{EXPECT}

While the integrated width for the inclusive transition $B \!\to\! X\!+\! 
\gamma $
can be affected by the intervention of New Physics, the shape of the
prompt photon spectrum is more robust against it, at least within
non-exotic scenarios. Two classes of effects govern the photon
spectrum: the lowest order process $b \to q+\gamma $ would produce a
monoenergetic line $E_\gamma = \frac{m_b^2-m_q^2}{2m_b} \simeq m_b/2$
characteristic of two-body kinematics; gluon bremsstrahlung adds
a continuous hadronic (partonic) mass spectrum all the way up
to $m_b$ thus producing a radiative tail covering the whole range of
smaller photon energies.

Gluon field modes with momenta of order $\Lam$ introduce additional 
nonperturbative
effects: most notably they induce a primordial
energy-momentum distribution of the $b$ quark inside 
the $B$ meson -- an effect that had been
introduced on phenomenological grounds more than 20 years ago
\cite{alip,ACM}. The emergence of this `Fermi motion' in
the OPE-based QCD heavy quark expansion was pointed out in Ref. \cite{prl}. 
The related Doppler-smearing 
further broadens the photon spectrum and populates also the gap
$\frac{m_b}{2} < E_\gamma < \frac{M_B}{2}$ distinguishing quark-level and 
hadronic kinematics. To leading order in $1/m_b$ the effect of the primordial 
distribution is given by the convolution
of the short-distance--generated parton spectrum ${\rm d}\Gamma^{\rm 
pert}/{\rm d}E$
with the heavy quark distribution function $F$:
\beq
\frac{{\rm d}\Gamma}{{\rm d}E_\gamma} = \int {\rm d}k_+ \; \frac{{\rm 
d}\Gamma^{\rm
pert}(E_\gamma\!-\!\frac{k_+}{2})}{{\rm d}E}
F(k_+) \;.
\label{12}
\eeq
The natural variable for $F$ is the light-cone
momentum $k_+\!=\!k_0\!-\!k_z$ carried by the heavy quark in the parton
description. In analogy with the deep inelastic scattering structure functions 
it is
often traded for a dimensionless variable
$x=\frac{k_+}{\La}$ with
$\La=\lim_{m_b\to\infty} (M_B\!-\!m_b)$. The function $F(x)$ has
support from $-\infty$ to $1$, and its moments are given by the
expectation values of local heavy quark operators:
\bea
\nonumber
\int_{-\infty}^{1} \!\!\! F(x)\; {\rm d} x = 1\, &,& \;\, \int_{-\infty}^{1}
\!\!\!x\, F(x)\; {\rm d} x = 0 \\ \int_{-\infty}^{1} \!\!\!x^2\, F(x)\; {\rm 
d} x
= \frac{1}{\La^2\!} \frac{\mu_\pi^2}{3}\, &,& \;\,
\int_{-\infty}^{1} \!\!\!x^3\, F(x)\; {\rm d} x = -\frac{1}{\La^3\!}
\frac{\rho_D^3}{3}, \label{14}
\eea
etc. ($\mu _{\pi}^2$, $\rho_D^3$ denote the expectation values of the kinetic 
energy
operator and the Darwin term, respectively). At large negative
$x$ this function must rapidly vanish. In actual QCD the
moments and likewise the distribution function itself depend on the 
normalization
point, and this dependence is actually far more
pronounced than in the usual leading-twist functions of light
hadrons. The exact separation between quantum modes that are
 short-distance and belong to the hard ``perturbative'' kernel, and those 
which are
considered soft and thus incorporated into the nonperturbative `primordial'
effects, is to some extent arbitrary and is set by fixing a
normalization scale $\mu$. In this way both elements on the right hand side of
Eq.~(\ref{12}) are $\mu$-dependent, while the physical spectrum on the 
left hand side is not.

Even leaving aside effects of higher order in $1/m_b$, predicting the
photon spectrum would require detailed knowledge of the function $F(x)$,
which is not quite realistic. Predictions \cite{bsg} 
were obtained assuming a reasonable functional form of $F(x)$ and fitting 
its parameters to
reproduce the known and estimated $B$ expectation values
in Eq.~(\ref{14}). Such a choice for $F(x)$ is not unique, and one might be 
concerned
that an essential uncertainty is thus introduced into the spectrum. It turns
out, however, that its gross 
features are relatively stable as long as the first three or four
moments are fixed.

This stability applies in particular to the lower part of the spectrum, 
especially
when one considers the fraction of events with photon energies below a value 
$E$:
\beq \Phi_{\gamma}(E) = \frac{1}{\Gamma (B \to X_s+\gamma)}
\int_0 ^{E} {\rm d} E_{\gamma} \frac{{\rm d}\Gamma}{{\rm d}E_{\gamma}} \; . 
\label{3}
\eeq
The fraction of decay events with $E_\gamma < 2\GeV$ theoretically
comes out about $12\%$, significantly larger than the $5\%$ 
CLEO's fit would yield
and above $8\%$ one would obtain literally using the central data points.
Yet, it is qualitatively evident that sufficiently low
parts of the spectrum are shaped by emission of relatively hard gluons
where the perturbative description is adequate and the intervention of
nonperturbative dynamics is insignificant. This was manifest in the
analysis of Ref.~\cite{bsg}.

Of course, the uncertainty in the distribution function $F(x)$ can to
some extent affect the fraction $\Phi_{\gamma}(2\GeV)$ if, say, $F(x)$ has
a long tail towards large negative $x$. It is usually missed, however,
that such a scenario would imply a rather large
value for the Darwin operator $\rho_D^3$ (and other higher-order ones) by
virtue of their relation to the moments in Eq.~(\ref{14}). This would not go
unnoticed in other applications of the heavy quark expansion. Yet since our 
direct
knowledge of such expectation values is at present still limited, we will not 
draw
conclusions here based on this observation. Instead we merely want to point 
out
that a broader distribution $F(x)$ yields additional smearing, which tends to
increase the low-$E_\gamma$ fraction of decay events. To quantify these 
intuitive
assertions we consider in more detail the purely perturbative component of the
spectrum $\frac{{\rm d}\Gamma^{\rm pert}}{{\rm d}E_\gamma}$.

\subsection{The perturbative spectrum}
\label{PERT}

Throughout this paper we assume the radiatively induced decay to be
generated by a single local operator
\beq
L_{\rm weak} = h\, \bar{s}\sigma_{\mu\nu}(1\!+\!\gamma_5)b \,
F^{\mu\nu}\;.
\label{n4}
\eeq
Normalized at the scale $m_b$, this term indeed yields the dominant
contribution to the $b\to s\!+\!\gamma$ decay rate in the Standard Model.
Other contributions to the transition amplitude generated at scales 
below $m_b$ give sizeable corrections, in particular the usual
four-fermion operator $\bar{s}b\, \bar{c}c$ with the low-scale
annihilation of the $c \bar c$ pair into photon and hadrons. However, this
mostly modifies the total rate, whereas the decay fraction we are
interested in is hardly changed.

Perturbative gluon effects in the $b\to s\!+\!\gamma$ spectrum contain some 
theoretical
complexities; for doubly logarithmic Sudakov corrections arise due to the
emission probability $\propto \frac{{\rm d}\omega}{\omega} \, 
\frac{{\rm d}k^2_\perp}{k^2_\perp}$,
where $\omega$ and $k_\perp$ are the gluon energy and transverse momentum,
respectively. The gluon coupling grows at small $k_{\perp}$, which on one hand
enhances corrections and on the other mandates isolating soft gluons from the
perturbative kernel. At the same time, introducing the
cutoff at any reasonable factorization scale eliminates $\log$s as a
large parameter in actual $B$ decays: with maximal $E_\gamma\!=\! m_b/2$
below $2.5\GeV$ the parameter $\ln {\!\frac{E_\gamma}{k}}$ can barely reach 
even
unity. Therefore, we should not expect significant uncertainties from
possible higher-order corrections.

The basic features of the perturbatively generated distribution become
manifest already in the model for the leading-log spectrum improved to
include running of $\alpha_s$, analyzed in Ref.~\cite{bsg}. One 
has\,\footnote{This ansatz in fact produces even a somewhat {\it harder} 
spectrum than the one from the
exact DL result based on soft factorization, which is evident when
double emission is considered. If the soft gluons emitted individually lower
$E_\gamma$ by amounts $\delta_1$ and $\delta_2$ with probabilities ${\rm d} 
w_1$ and ${\rm d} w_2$ respectively, double emission would
yield $E_\gamma=\frac{m_b}{2}\!-\!\delta_1\!-\!\delta_2$ with the probability
$\propto {\rm d} w_1\,{\rm d} w_2$, while the ansatz places this
contribution at $E_\gamma$ equal to the lower of the two energies,
$E_\gamma=\frac{m_b}{2}\!-\!{\rm max}\{\delta_1,\delta_2\}$.}
\beq
\frac{1}{\Gamma}\, \frac{{\rm d}\Gamma^{\rm pert}}{{\rm d} E_\gamma} =
- \frac{{\rm d}S(E_\gamma)}{{\rm d} E_\gamma}\,, \qquad S(E_\gamma)= 
e^{-w(E_\gamma)}
\label{20}
\eeq
where $S$ is the square of the Sudakov formfactor, and $w(E_\gamma)$
is the perturbative probability for a sufficiently hard gluon to be
emitted so that the photon is left with energy below $E_\gamma$. Using the
fact that soft gluons couple with $\alpha_s(k_\perp)$ and cutting the
integration at $k_\perp \!\le \!\mu$ where $\mu$ represents the separation
scale, the integrals are easily calculated \cite{bsg}:
$$
w(E_\gamma)\!=\! \frac{8}{3\pi}\left(\frac{2\pi}{9}\right)^2 \;\times
\mbox{\hspace*{111.7mm}}
$$
\beq
\mbox{\hspace*{8.7mm}}
\left\{
\begin{array}{ll}
\frac{1}{\alpha_s(m_b)} \ln{\frac{\alpha_s(\sqrt{\vep m_b})}{\alpha_s(m_b)}} -
\frac{1}{\alpha_s(\vep)} \ln{\frac{\alpha_s(\vep)}{\alpha_s(\sqrt{\vep
m_b})}} & \;\;\;\vep \!\ge\! \mu \\
\frac{1}{\alpha_s(m_b)} \ln{\frac{\alpha_s(\sqrt{\vep m_b})}{\alpha_s(m_b)}} -
\frac{1}{\alpha_s(\mu)} \ln{\frac{\alpha_s(\mu)}{\alpha_s(\sqrt{\vep 
m_b)}}} - \frac{9}{2\pi}
\ln{\frac{\mu}{\vep}}\left[1-
\ln{\frac{\alpha_s(\mu)}{\alpha_s(\sqrt{\vep m_b})}} \right]\;\,
&\frac{\mu^2}{m_b}\!\le\! \vep\! <\! \mu \\
\frac{1}{\alpha_s(m_b)} \ln{\frac{\alpha_s(\mu)}{\alpha_s(m_b)}}
-\frac{9}{2\pi} \ln{\frac{m_b}{\mu}}
& \;\;\;\vep \!\le\! \frac{\mu^2}{m_b}
\end{array}
\right. 
\label{24}
\eeq
\vspace*{-5mm}
$$
\vep = m_b-2E_\gamma\;.
$$
The representation of Eqs.~(\ref{20})--(\ref{24}) has the added advantage that 
the integrated
fraction $\Phi_{\gamma}(E)$ of the decay events with $E_\gamma \!<\! E$ is 
directly given by
$1\!-\!S(E)= 1\!-\!e^{-w(E)}$. The effective `soft' coupling 
in Eq.~(\ref{24}) is known to two
loops \cite{32}.

The perturbative photon spectrum following from
Eqs.~(\ref{20})-(\ref{24}) was discussed in Ref.~\cite{bsg}. The relevant 
point is
that $S(E)$ does not depend on
$\mu$ as long as $m_b\!-\!2E \!>\! \mu$. This means that this part of the
spectrum is shaped by gluons harder than $\mu$; in practice this scale
is around $0.9\GeV$ for $E\!=\!1.9\GeV$. Therefore, the bulk of the
perturbative spectrum below $2\GeV$ is indeed generated by
sufficiently hard gluons, and can be trusted. Since the
fraction $\Phi_\gamma (E)$ is still small, exponentiation of the soft
emissions does not produce a radical numerical change. It is important,
however, that this applies only as long as the cutoff in soft emissions
is implemented.

The upper part of the spectrum is mostly determined by the primordial
distribution function $F(x)$ with perturbative radiation affecting mainly the 
height
of the spectrum and somewhat widening its
shape. To a crude approximation, the central part of $F(x)$ can be
directly taken from the observed distribution at $E_\gamma \!>\! 2.2 \GeV$.

The lower part of the spectrum is to some extent affected by the
primordial Fermi motion as well, since $\Phi^{\rm pert}_\gamma (E)$
gets convoluted with $F(x)$:
\beq
\Phi_\gamma(E)= \int {\rm d}k_+ \,\Phi^{\rm pert}(E\!-\!k_+) \,F(k_+)\, \;.
\label{32}
\eeq 
First, a long tail of $F(x)$ to negative $x$ can populate the domain of lower
$E_\gamma$ even in the absence of hard bremsstrahlung at
all. This clearly would only enhance the estimate of $\Phi_\gamma(2\GeV)$. 
Secondly,
there is an effect of smearing due to typical $k_+ \sim \Lam$. 
Since the perturbative spectrum decreases sharply with increase
of $\frac{m_b}{2}\!-\!E_\gamma$, this normally enhances $\Phi_\gamma(E)$ as 
well, at
least with $E$ in the deep perturbative domain -- the perturbative
distribution is much wider than the intrinsic primordial
smearing in the $1/m_Q$ expansion. However, in practice typical
momenta in $F$ could be of similar scale as $\frac{m_b}{2}\!-\!E_\gamma$; when 
$k_+$
kicks up the effective perturbative energy $E_\gamma\!+\!k_+/2$ to the
domain where $\Phi^{\rm pert}_\gamma$ is nearly saturated, the above 
approximation
does not work anymore. Therefore, the primordial smearing can, in
principle, decrease the purely perturbative estimate -- in particular, when 
$\Phi^{\rm
pert}_\gamma(E)$ is already significant and $F(x)$ is a broad function.
In practice, the nonperturbative smearing only enhances the tail of the
spectrum in the relevant domain.

The double $\log$ approximation turns out to be far too crude for $B$ decays.
This can be assessed by comparing the exact ${\cal O}(\alpha_s)$ spectrum 
\cite{tree} with the one
obtained expanding the double $\log$ expression
in $\alpha_s$: the resulting spectrum exceeds the complete expression by
more than a factor of $3$. Moreover, keeping double and single
$\log$s alone would yield an enhancement -- rather than suppression --
factor in the Sudakov exponent for $E_\gamma$ as large as
$\frac{m_b}{2}\!-\! 75\MeV$. In this situation one better 
relies on the exact one loop spectrum:
\beq
\frac{1}{\Gamma}\, \frac{{\rm d}\Gamma^{\rm pert}}{{\rm d} E_\gamma} =
\frac{\alpha_s}{\pi}\,
\frac{(2x^2\!-\!3x\!-\!6)x+2(x^2\!-\!3)\ln{(1\!-\!x)}}{3(1\!-\!x)},\qquad
x=\frac{2E_\gamma}{m_b}\;.
\label{n10}
\eeq
It is clear, however, that using a fixed coupling $\alpha_s(m_b)$ would
significantly underestimate the actual spectrum. The BLM part of the
second-order corrections has been computed in Ref.~\cite{llmw} and was
indeed found to increase the spectrum by about $60\%$ for low enough
$E_\gamma$.

There are subtle physical reasons for this BLM approximation to be of
limited validity in the domain of relevant $E_\gamma$. The BLM
corrections for a given photon energy are obtained integrating over 
all gluon configurations consistent with the invariant mass of the
strange quark-gluon system. In particular, the gluon can be 
energetic and collinear, or relatively soft yet emitted at large angles:
\beq
M_X^2= 2m_b(m_b\!-\!E_\gamma) \approx \frac{k_\perp^2}{\omega}\,.
\label{n20}
\eeq
The scale of the strong coupling, however is driven mainly by
$k_\perp$. Hence, the effect of soft gluons is underestimated
when averaged with those from short-scale configurations.

This becomes important when double gluon emissions are considered. 
Their probability is proportional to $\alpha_s^2$, and they are missed by the
BLM terms. In contrast to ordinary corrections, this effect has an extra
enhancement factor of $\ln{(1-x)}/x$ for each emission. Thus the
common wisdom about the BLM dominance is not expected to apply here. And
there is an additional numerical enhancement: double (or any multiple)
emission allows lowering $E_\gamma$ by the same amount with softer
gluons (e.g.\ lower $k_\perp$ for same $\omega$); since $\alpha_s$ 
grows fast, this would yield a significant increase of the radiative tail 
overlooked by the naive BLM scale fixing. This can also be understood in a
complimentary way applying the OPE to soft perturbative bremsstrahlung.

This problem obviously does not arise if the Wilsonian prescription 
for separating long and short distance dynamics is
implemented when computing the perturbative spectrum to avoid double counting
of soft gluon modes. Then the infrared domain is excluded and the
coupling never grows too large. Moreover, soft gluons are simply excluded
from computations regardless of the strength of the coupling. 
This eliminates both the possible enhancement in the soft emission amplitudes
and strongly suppresses multiple emissions altogether; 
for the latter would now lower $E_\gamma$ by too large an amount. This 
approach allows using a single-gluon--generated spectrum and makes the 
BLM-type accounting for the running of $\alpha_s$ trustworthy.

The estimated fraction $\Phi_\gamma ^{\rm pert}(E)$  depends mainly on
the exact value of the $b$ quark mass.  Actually $\Phi_\gamma ^{\rm
pert}$ is primarily a function of $\frac{m_b}{2}\!-\!E$ rather than of
$E$ itself.  It is evident that one should use the perturbative running
low-scale mass $m_b(\mu)$ here, not the pole mass: excluding emissions
of soft gluons requires -- due to gauge invariance -- discarding 
analogous virtual corrections, including the soft corrections in the
$b$  quark  self-energy. Then the pole mass never appears in the
problem, and the  heavy quark is seen in perturbation theory as a
particle with the mass near  $m_b(\mu)$.

In principle, the Wilsonian prescription can be introduced in different
ways.  A consistent scheme suitable also for genuinely Minkowskian
processes  and beyond one loop has been elaborated \cite{blmope,five}. In
essentially  one-loop corrections (including arbitrary order in the BLM
resummation) it reduces to a cutoff in the gluon energy $\omega > \mu$
(in the perturbative computations) in the heavy quark restframe. The
value of the ``kinetic'' running quark mass and its running is well
known (for a review see Ref.~\cite{ioffe}):  
\beq 
m_b(1\GeV) = (4.57\pm 0.05)\GeV 
\label{n28} 
\eeq 
(a somewhat larger value -- $4.63\GeV$ -- however with twice the 
uncertainty was reported by CLEO \cite{CLEO1}). A subtlety should  
be kept in mind though: with such a
cutoff scheme there are different short-distance masses perturbatively
related to  each other. The one entering here is $m_0(\mu)$, the mass
defining the  rest energy, rather than the kinetic mass in
Eq.~(\ref{n28}). The former slightly exceeds $m_b(\mu)$: 
\beq
m_0(\mu) = m^{\rm kin}(\mu)+\frac{4}{9}\frac{\alpha_s}{\pi}\mu \left(1+
\frac{\beta_0\alpha_s}{2\pi}
\left[\ln{\frac{m_b}{2\mu}}\!+\!\frac{13}{6}\right] +...\right)
+{\cal O}\left(\alpha_s\frac{\mu^3}{4m_b^2}\right)\;.
\label{n30}
\eeq
The difference constitutes about $60\MeV$ at $\mu\!=\!1\GeV$. (The
straightforward all-order BLM summation yields a larger shift, however
for consistency we retain only terms through $\beta_0\alpha_s^2$.) 
With such $m_b$ the ratio $\frac{2E_\gamma}{m_b}$ is about 
$0.86$ at $E_\gamma\!=\!2\GeV$.

The perturbative expressions existing in the literature \cite{tree,llmw}
for  the one-loop spectrum do not include the effect of an infrared cut,
even though this computation is not difficult. For low enough $E_\gamma$
this would not affect the single-gluon spectrum as have been already
noted for the  double log corrections. Strictly speaking, a gluon with
$\omega \!\le\! \mu$ can bring  $E_\gamma$ as low as 
$\frac{m_b}{2}\!-\!\mu$, so
the cut affects the spectrum in a wider domain. This  is possible,
however, only when the gluon and the $s$ quark fly back-to-back, a
rather rare and -- being governed by a short-distance coupling --
suppressed  configuration. The enhanced kinematic sensitivity is absent
as soon as the gluon is emitted into the forward hemisphere. Hence, to a
good approximation the effect of the cutoff can be neglected at
$E_\gamma$ below $\frac{m_b-\mu}{2}$ as it happened in the double $\log$
ansatz (\ref{24}).

Above this borderline the perturbative spectrum is modified more 
significantly, its growth slowed
down with the spectrum vanishing 
around $\frac{m_b}{2}-\frac{\mu^2}{2m_b}$. It also has the
two-body $\delta$-function at $E_\gamma=\frac{m_b(\mu)}{2}$ suppressed by a 
finite amount by virtual corrections.

\begin{figure}[hhh]
\begin{center}
\vspace*{-3mm}
\mbox{\psfig{file=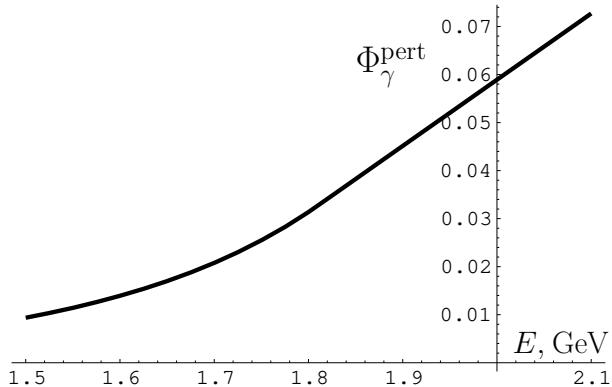,width=8cm}}
\end{center}
\vspace*{-7mm}
\caption{{\small 
Perturbative fraction of decay events $\Phi_\gamma^{\rm pert}(E)$ 
having $E_\gamma\!<\!E$.
} 
}
\end{figure}

The estimated fraction of the decay events generated solely by
perturbative gluons is shown in Fig.~2. We see that one expects
$\Phi^{\rm pert}_\gamma(2\GeV)$ to be about $6\%$.

\subsection{Combined spectrum}
\label{COMB}

The overall spectrum is obtained by convoluting the perturbative
contribution  with the light-cone distribution function $F(x)$. Hence it
requires the perturbative part in the whole domain including the
interval  $m_b-2\mu < 2E_\gamma \le m_b$ where it is modified by the
gluon cutoff $\mu$.  However, the exact details in this relatively
narrow domain are not too significant due to primordial smearing, as
long as the overall normalization, $\Phi^{\rm pert}_\gamma(E)=1$ at
$E>\frac{m_b}{2}$ is maintained. Using the double $\log$ computation as
a guide, we extrapolate  $\Phi^{\rm pert}_\gamma(E)$ linearly above
$E_0=\frac{m_b-\mu}{2}$; the  end-point $\delta$-function in the
spectrum is fixed by the normalization constraint. This approximation is
expected to slightly underestimate  $\Phi^{\rm pert}_\gamma$ in the
lower end of the window $\frac{m_b}{2}-\mu$,  and overestimate it near
the maximal $E_\gamma$ except the endpoint itself. The bias in the
resulting total fraction should then be insignificant and presumably
only underestimates $\Phi_\gamma(E)$ at $E$ near $2\GeV$.

As illustrated in Ref.~\cite{bsg}, while the purely perturbative
spectrum and $\Phi^{\rm pert}_\gamma(E_\gamma; \mu)$ depend strongly on
the choice of $\mu$, the overall $\Phi_\gamma(E)$ comes out fairly
insensitive to it. For the $\mu$ dependence of $\Phi^{\rm pert}_\gamma$
is strongly offset  by the renormalization of $m_b(\mu)$, i.e. by the
shift in the starting point of  the parton-level spectrum. Therefore we
can infer numerical estimates using a particular choice of the cutoff
scale, namely $\mu\!=\!1\GeV$ for  which the nonperturbative parameters
are routinely fixed.

\begin{figure}[hhh]
\begin{center}
\vspace*{-3mm}
\mbox{\psfig{file=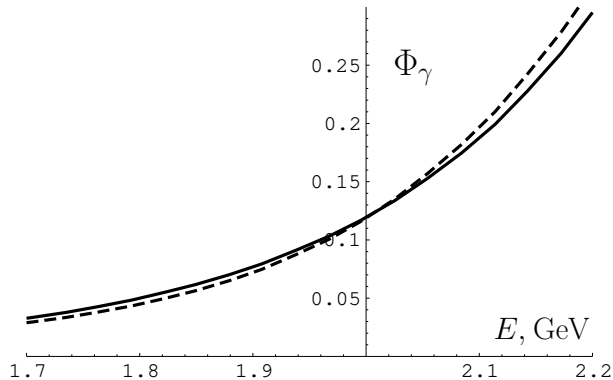,width=8cm}}
\end{center}
\vspace*{-7mm}
\caption{{\small 
Fraction of decay decay events $\Phi_\gamma(E)$ including 
both perturbative bremsstrahlung and primordial Fermi motion. Solid 
and dashed lines correspond to ans\"{a}tze Eqs.~(\ref{n32}) and 
(\ref{n36}) for the light-cone distribution function, respectively.
} 
}
\end{figure}

The expected fraction $\Phi_\gamma(E)$ is shown in Fig.~3 and the
spectrum itself in Fig.~4; we superimposed the experimental points onto
our predictions in the latter. We took 
the ansatz of Ref.~\cite{bsg} 
for the distribution function $F(x)$, but simplified it to an essentially
a single-parameter one:
\beq
F(k_+)=N\,(k_+\!-\!\La)^\alpha \,e^{-ck_+} \:\theta(k_+\!-\!\La)\;.
\label{n32}
\eeq
$c$ gauges the scale of the light-cone momentum, and $\alpha$ determines
the dimensionless ratio $\frac{\mu_\pi^2}{3\La^2}=\frac{1}{1+\alpha}$. 
Numerically $\alpha\!=\!2$ yields a value very close to the experimental
ones with $\mu_\pi^2(1\GeV)=0.43\GeV^2$ and $\La(1\GeV)=0.65\GeV$. It is
curious to note that ansatz (\ref{n32}) yields for the Darwin
expectation value 
\beq
\frac{\rho_D^3}{3}=\frac{2}{\La} \,\left(\frac{\mu_\pi^2}{3}\right)^2
\label{n34}
\eeq
for arbitrary $\alpha$. Such a value is favored by theory and would hold
exactly if the SV sum rules were saturated by the $P$-wave states at a
single mass.

\begin{figure}[hhh]
\begin{center}
\vspace*{-3mm}
\mbox{\psfig{file=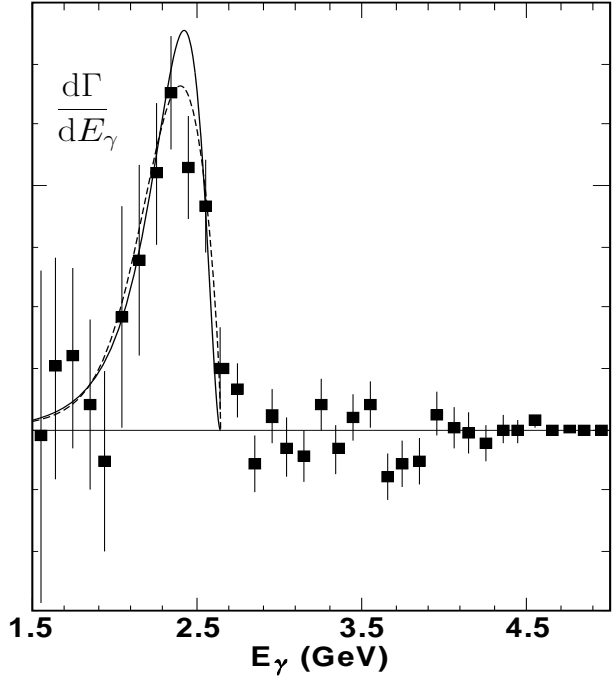,height=9cm,
width=8cm}}
\end{center}
\vspace*{-7mm}
\caption{{\small 
Expected photon spectrum for the two ans\"{a}tze for $F(k_+)$. The relative 
normalization of experimental points and theoretical predictions is not 
fixed. 
} 
}
\end{figure}

To check the sensitivity to the shape of $F$ we have also employed the
alternative ansatz 
\beq
\tilde F(k_+)= \tilde N\,(k_+\!-\!\La)^\beta \,e^{-(d(k_+\!-\!\La))^2}
\:\theta(k_+\!-\!\La)
\label{n36}
\eeq
with the tail fading out much faster, and $\beta$ adjusted as to
reproduce the same kinetic expectation value for a given $m_b$. (At
$\alpha$=2 this corresponds to $\beta\simeq 0.658$ and yields a factor
of $2/3$ smaller Darwin expectation value). As expected, 
$\Phi_{\gamma}(E)$ is  practically insensitive to the concrete ansatz; 
even the difference in the spectrum itself lies below the effects of
higher-order nonperturbative corrections. It should be noted that the
predictions shown in Fig.~4 do not include Doppler smearing due to the 
initial velocity of $B$ mesons produced  at $\Upsilon(4S)$. Assuming 
absence of a correlation of experimental measurements with the direction of the
decay relative to the $B$ momentum, the additional broadening is easily
incorporated. The corresponding plots are given in Fig.~5.

\begin{figure}[hhh]
\begin{center}
\vspace*{-3mm}
\mbox{\psfig{file=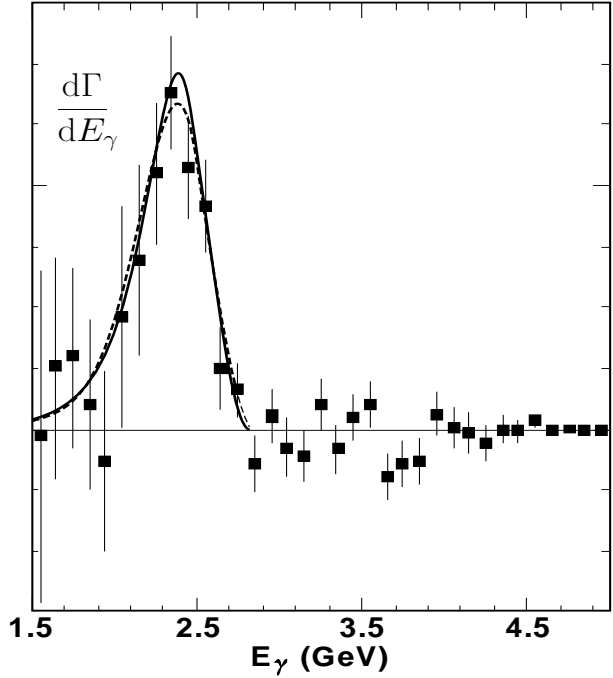,height=9cm,
width=8cm}}
\end{center}
\vspace*{-7mm}
\caption{{\small 
Photon spectrum in the lab frame where $\Upsilon(4S)$ are at rest, 
including Doppler smearing with $\beta_B\!=\!0.067$. The relative 
normalization is arbitrary.
} 
}
\end{figure}

Therefore, we arrive at the conclusion that about $12\% $ of decay
events should have $E_\gamma \!<\! 2\GeV$. Half of this probability
is generated perturbatively, and it already exceeds the $5\%$ of CLEO's
fit. The fact that CLEO's fit represents too narrow a spectrum can be 
inferred also from the size of the second moment or dispersion through
an alternative perspective: the value quoted by CLEO is essentially
oversaturated by  perturbative contributions alone.

It should be noted that although the fit was used in the analysis of
Ref.~\cite{CLEO1}, the total decay probability was obtained there
adopting $\Phi_\gamma(2\GeV)\!=\!8.5^{+5.5}_{-2.7}\%$ from
Ref.~\cite{KN}, even though the fit itself yielded about $5\%$.

At present the part of the spectrum below $2\GeV$ does not seem very
certain. If, however, future measurements establish that significantly
less than $10\%$  of the spectrum resides there, this 
would send the alarming message that the standard treatment of strong 
interaction effects -- in particular the gluon corrections as they are
routinely applied  to decay distributions for beauty hadrons -- is
unreliable. This in turn would cast  doubts on the numerical accuracy of
extracting $m_b$, $\mu_\pi^2$ etc.\ from the photon spectrum and even
more on relating the lepton endpoint spectrum in $b\to u\,  \ell \nu $
to the $b \to s \!+\!\gamma$ spectrum.

It can well be that the fully integrated spectrum is still robust
against such effects; however the higher moments and the shape in
general would be  affected. If CLEO's fit were to be vindicated by
future data, one could not extract the quantity $\bar \Lambda$ (let alone 
the kinetic expectation value) using the quoted formulae; for
those expressions were based on the assumption that gluon radiation can
be  described perturbatively over the whole range up to
$E_\gamma\!=\!2\GeV$.

The simplest resolution of a possible discrepancy between the CLEO spectrum and
the theoretical expectation is that in particular the data point at the
$1.9\GeV$  bin  represents a low fluctuation. Future data will settle
this issue.

Of course, the first and, in particular, second moment of the photon
energy  decrease if a spectrum closer to QCD expectations is used.
Unfortunately, one cannot re-evaluate them based only on the data points
given in Ref.~\cite{CLEO1}. Extracting the value of the $b$ quark mass
it relied only on the moments evaluated over the range $E_\gamma 
\!>\!2\GeV$,  hence there was a limited sensitivity of the purely
experimental input to the lower part of the spectrum. 
However, this simply relegates the sensitivity to actual physics of
gluon bremsstrahlung to theoretical evaluation of such defined moments.

\section{Skeptical remarks}
\label{SKEPT}
          
One of the practical applications of measuring the prompt photon
spectrum is  an attempt to estimate more reliably the fraction of
semileptonic $b\to u$ decays expected theoretically with $E_\ell$ in the
end-point region beyond the  kinematic range for $b\to c$. While
appealing as a nontrivial physical statement, in  practice the relation
between the integrated semileptonic end-point spectrum 
\beq
\frac{1}{\Gamma_{\rm sl}(b\to u)} \int_E^{\frac{M_B}{2}} {\rm d}E_\ell\,
\frac{{\rm d}\Gamma_{\rm sl}(b\to u)}{{\rm d}E_\ell}
\label{37}
\eeq 
and the corresponding weighted $B\to X_s\!+\!\gamma$ spectrum 
\beq
\frac{1}{\Gamma_{bs\gamma}} \int_E^{\frac{M_B}{2}} {\rm d}E_\gamma\,
\left(\frac{M_B}{2}-E_\gamma \right)
\frac{{\rm d}\Gamma_{bs\gamma}}{{\rm d}E_\gamma}
\label{38}
\eeq 
is rather fragile -- a fact emphasized already in the first dedicated 
paper \cite{motion}: the actual value of $m_b$ turns out to be
insufficiently large to make higher  order corrections irrelevant, and
the allowed interval of lepton energies $E_\ell > 
\frac{M_B^2-M_D^2}{2M_B}$ is too narrow to suppress higher-order
corrections violating universality of the Fermi motion. There are two
basic nonperturbative effects which are potentially dangerous here; we
illustrate  them in more detail than it was done in Ref.~\cite{motion}.

The first one is related to a particular class of $1/m_b^3$  corrections
associated with four-fermion operators responsible for generalized  Weak
Annihilation (WA) effects. While typically at a few percent level in
fully integrated $B$ widths, they are concentrated in the  end-point
domain in semileptonic decays and thus are enhanced by an order of 
magnitude for the narrow slice in $E_\ell$ \cite{WA}. Their size is
controlled  by the poorly known nonfactorizable terms in the expectation
values  $\matel{B}{\bar{b}\gamma_\mu(1\!-\!\gamma_5)u
\bar{u}\gamma_\nu(1\!-\!\gamma_5)b}{B}$; they are different in  charged
and neutral $B$ and in general do not vanish even in $B^0$: 
\beq 
\delta \Gamma_{\rm sl}(b\to u) \simeq 
\Gamma^0_{\rm sl}(b\to u) \: \frac{32\pi^2}{m_b^3}\,
\frac{\matel{B}{\bar{b}\gamma_k(1\!-\!\gamma_5)u\,
\bar{u}\gamma^k(1\!-\!\gamma_5)b}{B} }{2M_B} \,. 
\label{40} 
\eeq 
The nonfactorizable contributions were addressed in  Ref.~\cite{four};
the non-valence operators, however, are rather uncertain. The reasonable
estimate of Ref.~\cite{vub} suggests that this  effect constitutes
$10\div 30\%$ of the semileptonic width with $E_\ell \!>\!  2.3\GeV$. A
more detailed analysis can be found in the original papers 
\cite{WA,vub}.

No such effect arises for the $B\!\to\! X_s\!+\!\gamma $ width at tree
level. It  is generated to order $\alpha_s$ \cite{motion} and is given
by a different  four-fermion operator. The correction to $\Gamma (B
\!\to\! X+\gamma )$ is expected  smaller here; a more careful
consideration reveals that the end-point fraction can be affected, yet
less significantly than in $B \!\to\! X_u \,\ell \nu$.

The second problem is of a general origin and related to the fact  that
the interval in the electron (or photon) energy $E_\ell > 2.3\GeV$  is
too narrow in practice \cite{motion}. To quantify this observation, let
us recall the $1/m_Q^2$ corrections in the $b\to  u$ lepton spectrum
\cite{dpf,prl}:
$$
\frac{1}{\Gamma_0}\frac{{\rm d}\Gamma_{\rm \!sl}}{{\rm d}E_\ell} 
\!=\! 2 \!\left(\frac{\!2E_\ell\!}{m_b}\right)^{\!\!2}\!
\left\{ 3\!-\!\frac{4E_\ell}{m_b}\!-\!\left[\frac{10
E_\ell}{m_b^3}\!+\!\frac{2}{3m_b}\delta(\mbox{$\frac{m_b}{2}$}\!-\!E_{\ell\!}
)
\!-\! \frac{E_\ell^2}{3m_b^2}(m_b\!-\!E_{\ell\!} )
\delta'(\mbox{$\frac{m_b}{2}$}\!-\!E_\ell)_{\!}
\right]\!\mu_\pi^2 + \right.
$$
\beq
\left.
\left[
\frac{2}{m_b^2}+\frac{10E_\ell}{3m_b^3}-
\frac{11}{12m_b}\delta(\mbox{$\frac{m_b}{2}$}\!-\!E_\ell)
\right]\mu_G^2
\right\}
\,,
\label{50}
\eeq
and examine the effect of the {\it chromomagnetic} operator. It has  a
negative $\delta$-function contribution in the end point:
\beq
\frac{1}{\Gamma_0}\,
\frac{{\rm d}\Gamma_{\rm sl}}{{\rm d}E_\ell}
\raisebox{-.4em}{$\left.\rule{0mm}{1.4em}\right|_{E_\ell \simeq
\frac{m_b}{2}}$ }
\!\!\!\!\simeq\;
2\left\{1 - \frac{11}{12}\,\frac{\mu_G^2}{m_b}\, 
\delta(\mbox{$\frac{m_b}{2}$}\!-\!E_\ell)\right\}\;+\;
\mbox{term}\,\propto \mu_\pi^2
\,.
\label{52}
\eeq 

The spectrum can be neither singular nor negative, of course.  What this
means is that one has to average the lepton energy at the very least 
over an interval $\Delta_{\rm min} \!=\!
\frac{11}{12}\,\frac{\mu_G^2}{m_b}\simeq  70\MeV$; 
a similar shift and a need for smearing applies {\it a priori} to 
$b\!\to\!s\!+\!\gamma$ events as well. On the other  hand it is
quite evident that  this effect is completely missing from the Fermi
motion and from the  distribution function $F(x)$ -- in contrast to the
latter it is a non-universal  process-dependent nonperturbative
correction and would even have the opposite sign in decays of 
$B^*$.\footnote{Chromomagnetic field $\vec{B}(0)$ emerges as a
commutator of different spacelike  momentum operators of $b$ quark,
while Fermi motion depends only on a single  space momentum and energy
operator. The term $\propto \mu_\pi^2$ is the first term  in the effect
of the primordial motion.} Thus it is absent from the naive  leading in
$1/m_b$ relations between the observables in Eqs.~(\ref{37}) and
(\ref{38}).

The shift by $70 \MeV$ does not affect much the total decay rate -- yet 
it is very significant for the narrow end-point slice: for the parton
spectrum it would decrease the rate by a factor $1\!-\!\frac{11}{6}
\frac{\mu_G^2}{m_b(m_b\!-\!2E)}$. This is more than a $50$ percent reduction
for $E\!=\!2.3 \GeV$ even with conservative assumptions about $m_b$.

We can suggest a simplified way to improve the end-point relation 
between the photon and lepton spectrum close in spirit to the proposal
to  be made below in Sect.~5.2 for the invariant mass $M_X^2$
distribution. To account  for the chromomagnetic interaction effect in
Eq.~(\ref{52}) one can make a  shift in the lepton vs.\ photon energy by
a formally ${\cal O}(1/m_b)$ amount $\delta  E_\ell = \frac{2}{3}
\frac{\mu_G^2}{m_b}$: 
\beq
\frac{1}{\Gamma_{\rm sl}(b\to u)} \int_E^{\frac{M_B}{2}} {\rm d}E_\ell\,
\frac{{\rm d}\Gamma_{\rm sl}(b\to u)}{{\rm d}E_\ell}
\propto
\frac{1}{\Gamma_{bs\gamma}} \int_{E+\delta E_\ell}^{\frac{M_B}{2}} 
{\rm d}E_\gamma\, \left(\frac{M_B}{2}-E_\gamma \right)
\frac{{\rm d}\Gamma_{bs\gamma}}{{\rm d}E_\gamma}
\;.
\label{54}
\eeq
Physically this is motivated by the different interaction of fast 
light-like light quarks with the chromomagnetic field in the two
processes  leading to a relative energy shift. However this is not a
rigorous prescription and its  accuracy is difficult to quantify. Since
the effect is significant, it may  introduce an essential uncertainty
even with this improvement.

It is important to keep in mind that the above constraint on the 
minimal resolution is of a rather fundamental nature which goes beyond a
conceivable improvement of the Fermi motion description \cite{bsg}.
Knowledge of the subleading distribution functions would not enable  one
to shrink this interval. Of course, the literal expression for
higher-order terms like Eqs.~(\ref{40}), (\ref{52}) or (\ref{54}) going
beyond the  leading-twist expansion in this end-point domain cannot be
used literally to  improve the naive leading-order expressions, since
they are superimposed onto the nonperturbative leading-twist Fermi
motion. In principle, it cannot be excluded that the latter  somewhat
softens the numerical instability of the end-point expansion -- yet  one
cannot count on a radical improvement of the accuracy: the fact that 
allowing only a $70\MeV$ interval of the end point energy yields a
$100\%$  error in the spectrum in the best scenario, is a benchmark
which is difficult to  get around.

From the general perspective the problem is related to presence of 
another parameter in the problem -- the ratio of the scale
$\frac{m_c^2}{m_b}$  to the typical hadronic mass. With this scale
amounting to only a few times  $\frac{\mu_G^2}{m_b}$ the relation for
the end-point spectrum is violated by a  fixed amount even for
asymptotic values of $m_b$. The accuracy of the relation is not
governed solely by $1/m_b$ as is usually stated,  and the actual numbers
favor a possibility for significant corrections. This  point is often
missed in the literature.

 \section{Improving the $M_X^2$ distribution in $B \to X_u \,\ell \nu $ }
 \label{NEW}

The integrated width $\Gamma (B \to X_u \,\ell \nu )$ is under good 
theoretical control \cite{vadem}. Yet to measure it one has to
disentangle it from the dominant $B \to X_c \,\ell \nu $ width. Vertex
detection  is not so highly efficient that one could achieve this goal
solely or even  mainly by rejecting the $b \to c$ background through
finding the secondary charm decays.  One has to apply cuts of a mainly
kinematic nature. The most direct way  is to consider the hadronic
recoil mass spectrum $\frac{{\rm d}\Gamma}{{\rm  d}M_X}$ and impose a
cut in the invariant hadronic mass $M_X \!<\! M_D$, which  has the
advantage of retaining almost all $b \!\to\! u$ events. For experimental
reasons one will have to lower the cut to confidently  reject the
$b\!\to\! c$ background. The situation is quite favorable since  even a
cut as low as $1.5\GeV$ leaves more than half of the signal events. Yet
one still wants to place it as high as possible, preferably  above
$1.65$ or even $1.7\GeV$. Considering the huge statistics that can be 
accumulated at modern $B$ factories the central issue becomes how well
one can  theoretically evaluate the fraction of the $b \to u$ width
falling into a restricted domain of phase space. Raising the cut reduces 
the rejected fraction and thus suppresses the theoretical uncertainty in
the  measured rate even without precise control over the fraction.

While the $M_X$-spectrum is sensitive to the values of the 
nonperturbative parameters as well as the heavy quark distribution
function $F(x)$ describing Fermi motion, the dependence on the latter is 
moderate if the current knowledge and the constraints on $F(x)$ are
incorporated in  full. This particularly applies to the partially
integrated spectrum.  Following Refs.~\cite{keymx} we consider the
fraction of events with the  hadronic recoil mass below $M_{\rm max}$: 
\beq 
\Phi_{\rm sl} (M) = \frac{1}{\Gamma_{\rm sl} (b\to u)}\, 
\int _0^{M}\! {\rm d}M_X \, 
\frac{{\rm d}\Gamma_{\rm sl}(b\to u)}{{\rm d}M_X}
\label{60}
\eeq 
Even without a dedicated analysis of all the uncertainties it  is
evident that $\Phi_{\rm \!sl\!}(1.7\GeV)$ must exceed $0.55$ and be 
below $0.9$. This observable is thus known with at least $\pm 24\%$ 
accuracy translating into a $\pm 12\%$ error bar in $|V_{ub}|$. Even 
allowing for additional stretching of uncertainties this would represent
at the  moment a good benchmark for quantities relevant for $|V_{ub}|$.
We found no  substance to claims of unrecoverably large uncertainties in
this observable.

Ultimately we would like to be able to decrease the theoretical 
uncertainties in $\Phi_{\rm sl}(M)$ further still -- and achieve it
without  having to push the cutoff $M$ too close to $M_D$ or degrading
the confidence in the assessed error bars. One obvious  possibility is
to introduce an additional cut rejecting small invariant mass  $q^2$
($q$ denotes the momentum of the lepton pair) where the  strong
interaction effects on $M_X^2$ are maximal due to the significant 
recoil.\footnote{A cut, say on the lepton pair space momentum 
$|\vec{q}\,|$ achieves a similar purpose since it is kinematically
related to  $M_X^2$ and $q^2$.} Choosing, for instance $q^2 \!\ge\! 0.35
m_b^2$  practically eliminates the impact of the primordial Fermi motion
on $M_X \!<  \!1.7\GeV$ events. However, imposing additional cuts has
the serious drawback of introducing a significant dependence on
higher-order effects  which are difficult to control and which are all
too often missed in estimates. The overall energy scale governing
the intrinsic `hardness' of the  decay rate deteriorates being driven at
large $q^2$ by $m_b\!-\!\sqrt{q^2}$  instead of $m_b$ itself
\cite{five}. Fortunately, there is an alternative way not employing
additional cuts.

The idea is based on the realization that the dangerous small-$q^2$ 
domain in $B\!\to\!X_u\,\ell \nu $ decays is nearly identical to the
case of  $B\!\to\! X_s\!+\!\gamma$ in respect to the hadronic mass
distribution, at least in the limit of large $m_b$. Moreover, its
accuracy is governed by a genuinely $1/m_b$ expansion. Therefore, we can
directly utilize experimental  information about the $B\!\to\! X_s
\!+\!\gamma $ spectrum to control this kinematics, in a  far more
reliable way than the charged lepton end-point spectrum.

\subsection{Scaling behavior and universality relations}

To illustrate the idea, let us consider first $B\!\to\!X_u\,\ell\nu $
at $q^2\!=\!0$. The concrete Lorentz structure of the decay vertex
($V\!-\!A$ in $b\!\to\! u\, \ell\nu$ and tensor in $b  \!\to\! s
\!+\!\gamma$) does not matter in the heavy quark limit; the effects of
strong interactions rather depend  on kinematics which are equal for
vanishing invariant mass of the  particles recoiling against the final
state hadronic system. The decay  distributions are governed by the
light quark propagator emitted with light-like  momentum $|k_u|\simeq
\frac{m_b}{2} \left(1\!-\!\frac{q^2}{m_b^2}\right)$.  Then one has
\beq
\frac{1}{\Gamma_{\rm sl}(q^2\!=\!0)}\, \frac{{\rm d} \Gamma_{\rm 
sl}(q^2\!=\!0)}{{\rm d} M_X^2} (M_X^2) \,=\, \frac{1}{2M_B}\:
\frac{1}{\Gamma_{bs\gamma}} \frac{{\rm d} \Gamma_{bs\gamma}}{{\rm 
d} E_\gamma}
\left( \mbox{$\frac{M_B^2\!-\!M_X^2}{2M_B}$}\right)\;,
\label{78}
\eeq
since kinematically $E_\gamma\!=\!\frac{M_B^2\!-\!M_{\rm 
hadr}^2}{2M_B}$. Here and in what follows we assume the $b\!\to\! u$ 
transitions when refer to semileptonic decays. Thus, at least for $q^2
\!=\! 0$ we can  directly obtain $1\!-\!\Phi_{\rm sl}(M)$ by integrating the
photon spectrum in  $B\to X_s\!+\!\gamma$ up to
$E_\gamma\!=\!\frac{M_B^2\!-\!M^2}{2M_B}$.

Relation (\ref{78}) does not hold when $q^2$ in the two processes  are
different, namely away from the point $q^2\!=\!0$ in the semileptonic 
decays. However, there is a scaling over the variable $M_{\rm
hadr}^2/\left(1\!-\!\frac{q^2}{m_b^2}\right)$ reflecting  the
universality of the Fermi motion \cite{motion}. It simply states that
the hadronic mass squared is proportional to the decay light quark
momentum $|k_u|$:
\beq
\frac{1}{\Gamma_{\rm sl}(q^2)}\, \frac{{\rm d} \Gamma_{\rm 
sl}(q^2)}{{\rm d} M_X^2}(M_X^2) \, =\,
\frac{1}{2M_B}\, \frac{m_b^2}{m_b^2\!-\!q^2}\:
\frac{1}{\Gamma_{bs\gamma}} \frac{{\rm d} \Gamma_{bs\gamma}}{{\rm 
d} E_\gamma}
\left( \mbox{$\frac{M_B}{2}\!-\!
\frac{M_X^2}{2M_B}\frac{m_b^2}{m_b^2\!-\!q^2}$}
\right)\;.
\label{80}
\eeq
This relation deteriorates for larger $q^2$ and breaks down when
$m_b^2\!-\!q^2$ becomes too low. However, at $q^2$ constituting a
significant fraction of $M_B^2$ the effect of Fermi motion on the
integrated fraction $\Phi(M)$ disappears which simply becomes unity in
this approximation.

Having this in mind, we can implement the scaling relation in  practice
to evaluate $\Phi_{\rm sl}(M)$. For the required rate with $M_X$ 
exceeding $M$ we have
\beq
\Phi(M)= \int {\rm d}q^2 \:
\frac{1}{\Gamma_{\rm sl}}\, \frac{{\rm d} \Gamma_{\rm sl}}{{\rm d} 
q^2} \, \int_{M^2} {\rm d}M_X^2 \;
\frac{1}{\Gamma_{\rm sl}(q^2)} \,
\frac{{\rm d}^2 \Gamma_{\rm sl}(M_X^2; q^2)}{{\rm d}M_X^2}
\;,
\label{82}
\eeq

with the last differential width in the scaling approximation given  by
Eq.~(\ref{80}). ($\Gamma_{\rm sl}(q^2)$ and $\frac{{\rm d} \Gamma_{\rm 
sl}}{{\rm d} q^2}$ denote the same total fixed-$q^2$ width.) Then we
have
\beq
1-\Phi_{\rm sl}(M; \chi)= \int_0^{M_B^2\!-\!M^2} \!\!\!
{\rm d}q^2 \, \chi(q^2)
\frac{1}{\Gamma_{\rm sl}}\, \frac{{\rm d} \Gamma_{\rm sl}}{{\rm d} 
q^2} \cdot \int_{0}^{\mbox{$\frac{M_B^2-\frac{M^2
m_b^2}{m_b^2\!-\!q^2}}{2M_B}$}}
\!\!{\rm d}E_\gamma \:
\frac{1}{\Gamma_{bs\gamma}}
\frac{{\rm d}\Gamma_{bs\gamma}}{{\rm d}E_\gamma}
\;,
\label{84}
\eeq
where $\chi(q^2)$ is a weight function one may (but does not have  to)
introduce to suppress contributions from small $q^2$ kinematics. 
Straightforward arithmetics yield
\beq
1-\Phi_{\rm sl}(M;\chi)= \int_0^{\mbox{$\frac{M_B}{2}\!-\!
\frac{M^2}{2M_B}$}} {\rm d}E_\gamma\;
\phi(E_\gamma,M;\chi)\,
\frac{1}{\Gamma_{bs\gamma}}
\,
\frac{{\rm d}\Gamma_{bs\gamma}}{{\rm d}E_\gamma}
\;,
\label{90g}
\eeq
with
$$
\phi(E_\gamma,M;\chi) = \int_0^{m_b^2-\frac{M^2 
m_b^2}{M_B^2-2E_\gamma M_B}} \!{\rm d} q^2\; \chi(q^2)\,
\frac{1}{\Gamma_{\rm sl}}\, \frac{{\rm d} \Gamma_{\rm sl}}{{\rm d} 
q^2}
$$
For not too large $q^2$ relevant in Eq.~(\ref{84}) the parton 
approximation 
\beq
\frac{1}{\Gamma_{\rm sl}}\, m_b^2 \frac{{\rm d} \Gamma_{\rm 
sl}}{{\rm d} q^2} = 6 \left(1\!-\!\frac{q^2}{m_b^2}\right)^2 - 4
\left(1\!-\!\frac{q^2}{m_b^2}\right)^3
\label{86}
\eeq
is quite accurate (see Ref.~\cite{keymx}). In what follows we  assume
$\chi(q^2)\!=\!1$ (no additional discrimination at all); then we arrive 
at the following {\it universality relation}
\beq
1-\Phi_{\rm sl}(M)= \int_0^{\mbox{$\frac{M_B}{2}\!-\!
\frac{M^2}{2M_B}$}} {\rm d}E_\gamma\;
\phi(E_\gamma,M)
\frac{1}{\Gamma_{bs\gamma}}
\,
\frac{{\rm d}\Gamma_{bs\gamma}}{{\rm d}E_\gamma}
\;,
\label{90}
\eeq
$$
\phi(E_\gamma,M) = 1-\frac{2 r^3}{(1\!-\!y)^3} + \frac{r^4}{(1\!-\!y)^4}
\,, \qquad y=\frac{2E_\gamma}{M_B}\,, \qquad r=\frac{M^2}{M_B^2}\;.
$$
It allows one -- as promised -- to express 
$\Phi_{\rm sl}(M)$ directly via the photon spectrum.

The weight functions $\phi(E_\gamma)$ depend on the chosen  $M_X$-cutoff
$M$ and vary from $1$ at $E_\gamma\ll M_B/2$ down to zero at the 
maximal $E_\gamma=\frac{M_B^2-M^2}{2M_B}$ where the single point 
$q^2\!=\!0$ contributes. They are shown in Fig.~6 for $M\!=\!1.6\GeV$,
$M\!=\!1.7\GeV$ and $M\!=\!1.8\GeV$. As anticipated, at $E_\gamma  \!<\!
2\GeV$ the weight $\phi(E_\gamma)$ is practically unity for all
reasonable cut masses $M$. This means that the low part of the photon
spectrum contributes only as a total integral, in other words, enters as
the overall fraction of events discussed in the previous sections. The 
spectrum at $E_\gamma \gsim 2\GeV$ enters with decreasing weight:
\beq
1-\Phi_{\rm sl}(M)\simeq \Phi_\gamma(2\GeV) + \int_{2\,{\rm 
GeV}}^{\mbox{$\frac{M_B}{2}\!-\! \frac{M^2}{2M_B}$}} {\rm d}E_\gamma\;
\phi(E_\gamma;M)
\frac{1}{\Gamma_{bs\gamma}}
\,
\frac{{\rm d}\Gamma_{bs\gamma}}{{\rm d}E_\gamma}
\;.
\label{92}
\eeq
The practical importance of clarifying the actual value of 
$\Phi_\gamma(2\GeV)$ becomes then manifest.

\begin{figure}[hhh]
\begin{center}
\vspace*{-3mm}
\mbox{\psfig{file=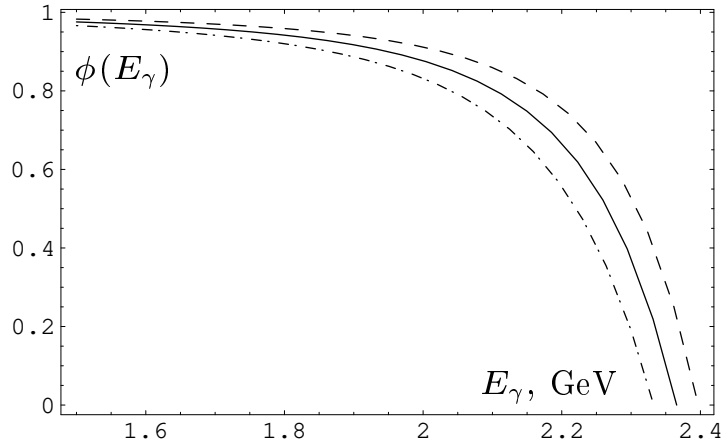,height=6cm}}
\end{center}
\vspace*{-8mm}
\caption{{\small 
Weight functions $\phi(E_\gamma)$ for three 
values of the cut on hadronic invariant mass, $M\!=\!1.6\GeV$ (dashed),
$M\!=\!1.7\GeV$ (solid) and $M\!=\!1.8\GeV$ (dot-dashed).} 
}
\end{figure}

The consequences of the Fermi motion description of the  large-recoil
light-cone kinematics similar to explicit relations (\ref{82}) have been
discussed in the literature more than once (see, e.g. the  recent paper
\cite{bauer}). Taking a different perspective, though, it is  usually
claimed that the outcome for $\Phi_{\rm sl}$ suffers from significant 
uncertainties. We do not share this pessimistic viewpoint for actual $B$
decays. For  example, the obvious constraint that $\Phi_{\rm sl}(M)$ is
exactly unity  regardless of dynamics at sufficiently high $M$
approaching $M_B$ is usually  neglected. As illustrated earlier in this
section, in reality we need to evaluate  only the {\it ab initio}
suppressed fraction $1\!-\!\Phi_{\rm sl}(M)$, and therefore even its
crude estimate does not introduce a significant uncertainty in
$|V_{ub}|$.

Universality relations and the scaling relations in general are clearly
not exact. In the given form they are robust against perturbative
effects; in particular  the Sudakov double $\log$ corrections would not
violate them. Yet they hold only to leading order in $1/m_b$. For
example, terms of order  $\La^2$ were neglected in $M_X^2$ compared to
the leading ones $\sim m_b \La$.  Since $1\!-\!\Phi_{\rm sl}(M)$ itself 
is a small fraction and thus introduces  only a correction in extracting
$|V_{ub}|$, its approximate knowledge already  suffices for our purpose.
Nevertheless further improvement can be achieved by  evaluating $1/m_b$
corrections to the universality relations Eqs.~(\ref{90}),  (\ref{92}).

\subsection{Leading $1/m_b$ corrections}

Accounting for the $\La^2$ terms referred to above is quite simple,  and
in general they can easily be incorporated in both semileptonic and 
radiative decays. However, this would not give the full answer. The
universality  relations make use of two basic facts, namely the
process-independence of the  leading-twist Fermi motion in the
light-cone kinematics and the scaling of  $M_X^2$ (or a certain function
of it) with $1\!-\!\frac{q^2}{m_b^2}$. While it is easy to  account for
the $1/m_b$ corrections to the former, $1/m_b$ scaling violation  is
governed by independent distribution functions.

There are two circumstances which help to overcome these obstacles.  In
practice, only a limited range of $q^2$ yields the unwanted background
due to Fermi motion; therefore the scaling violation in passing to $q^2
\! >\! 0$ is not dramatic and can be dealt with. Besides, we do not aim
at detailed distributions, but rather need an integrated probability. In
addition, in the small-$q^2$ domain where the Fermi motion is
potentially most important, its universality is ensured by the full $b$
quark mass rather than a  lower momentum scale. Below a way is suggested
to implement these ideas  in practice.

Instead of assuming the leading twist scaling in 
$M_X^2/(1\!-\!\frac{q^2}{m_b^2})$ for decay distributions, we can employ
its  $1/m_b$ improved ansatz:
\beq
\frac{1}{\Gamma_{\rm sl}(q^2)}\, \frac{{\rm d} \Gamma_{\rm 
sl}(\mu^2;q^2)}{{\rm d} \mu^2} \:=\: \frac{1}{2M_B}\,\frac{m_b^2-q^2}{m_b^2}\,
\,\frac{1}{\Gamma_{bs\gamma}} \frac{{\rm d} \Gamma_{bs\gamma}}{{\rm 
d} E_\gamma}
\left(\mbox{$\frac{M_B}{2}\!-\!\frac{\mu^2}{2M_B}\frac{m_b^2}{m_b^2-q^2}$})
\right)
\label{94}
\eeq
with $\mu^2$ given by \beq \mu^2=M_X^2+ A(\mbox{$\frac{q^2}{m_b^2}$} ) + 
B\mbox{$\left(\frac{q^2}{m_b^2}\right)$} M_X^2 + 
C\left(\mbox{$\frac{q^2}{m_b^2}$} \right)\frac{M_X^4}{M_B^2} \label{96}
\eeq where $A\sim \Lam^2$, $B\sim \frac{\Lam}{m_b}$ and $C\sim {\cal 
O}(1)$ are effective parameters accounting for $1/m_b$ corrections. In
other words, we assume $\mu^2$ instead of $M_X^2$ itself to scale
proportional to $1\!-\!\frac{q^2}{m_b^2}$.

In the simplest parton model one has, as an example,
\beq
A=\La^2\,, \qquad B=\frac{\La}{M_B}\,\frac{1+2x}{1-x}\,, \qquad
C=\frac{x}{(1\!-\!x)^2}\,; \qquad\; x=\frac{q^2}{M_B^2}\; , 
\label{98}
\eeq
which would incorporate kinematic effects. This is analogous to  deep
inelastic lepton-nucleon scattering, where the introduction of Nachtmann
variable leads to an extension of the scaling regime.  Simultaneously,
this can also account for corrections to the heavy quark symmetry for
different weak decay  currents ($V\!-\!A$ in $b\to u\,\ell\nu$ vs.\
tensor in $b\to s\!+\!\gamma$).  We do not rely on a model here, but
rather can fix the parameters theoretically  minimizing the strong
interaction corrections computed at given $q^2$ in the  standard $1/m_b$
expansion \cite{dpf,prl}. This determines the ansatz parameters in  a
model-independent way. After that, the straightforward integration 
(\ref{82}) would yield the $1/m_b$ and perturbatively improved version
of the universality  relation (\ref{90}).

A clarifying note is in order here. The standard $1/m_b$ expansion
yields nonperturbative effects only as an expansion around the free
quark kinematics, e.g.\ $M_X^2\!=\! (M_B\!-\!m_b)^2 \!+\!
(M_B\!-\!m_b)m_b(1\!-\!\frac{q^2}{m_b^2})$ for $b\!\to\!  u\,\ell\nu\,$
consisting of the $\delta$-function and its higher derivatives. Their
direct integration in the observables of interest makes little sense in
the present context (likewise, it cannot be used to quantify the
accuracy of various constrained inclusive probabilities in the case of
limited energy release, as can be sometimes seen in the literature).
Instead, here  we can compute at given $q^2$ the hadronic moments
($\aver{M_X^2}$, $\aver{M_X^4}$, \ldots) in terms of heavy quark
parameters $m_b$, $\mu_\pi^2$, $\mu_G^2$, $\rho_D^3$, \ldots directly
and in the  ansatz (\ref{94})-(\ref{96}), and require them to match. In
this way $1/m_b$  corrections are  properly accounted for.

\section{Conclusions and outlook}
\label{CONCL}

By the turn of the third Millennium $|V_{ub}|$ had been determined 
through a measurement of the inclusive rate $\Gamma (B \!\to
X_u\,\ell\nu)$.
Most  recently -- after this paper had largely been completed -- a new
result from  CLEO has appeared \cite{CLEOep} with an extraction of
$|V_{ub}|$ from the end  point lepton spectrum. Yet the theoretical 
reliability level of these methods varies considerably. The inclusive 
rate $\Gamma (B\!\to\! X_u \,\ell \nu )$ suffers  from little model
dependence. The main problem arises in distinguishing it against the
dominant $b \!\to\! c$ background and how to model the latter 
\cite{battag}.

In our considered judgement measuring the hadronic recoil mass  spectrum
in $B\!\to\! X_u \,\ell \nu $ will provide one with the most reliable 
extraction. Placing a high premium on model-independence we think that a
measurement of the lepton  energy endpoint spectrum {\it per se} cannot
provide a competitive extraction  of $|V_{ub}|$ as long as a cut around
$E_\ell \simeq \frac{M_B^2\!-\!M_D^2}{2M_B}$  remains, whether or not
the $B\to X_s\!+\!\gamma$ spectrum is known.

Nevertheless studying the end point lepton spectrum is of high 
theoretical interest. The fact that the analysis of Ref.~\cite{CLEOep} 
yielded a value for  $|V_{ub}|$ very close to other determinations 
places bounds on the
expectation values of four-quark operators. Naive estimates result in a
typical limit around 
\beq
\left|\,\frac{\matel{B}{\bar{b}\gamma_i(1\!-\!\gamma_5)u\, 
\bar{u}\gamma_i(1\!-\!\gamma_5)b}{B} }{2M_B}\,\right| < 0.012\GeV^3
\label{104}
\eeq
(an average over charged and neutral $B$ is implied) which lies in  the
range of previous estimates \cite{WA,Ds,four,vub}. A more careful
analysis  is required, though, to obtain reliable numbers. In
particular, an improvement in treating the end-point relation between
the  semileptonic and radiative spectra may turn the bound into a
nontrivial  measurement of these expectation values. The interpretation
would be much better  supported if measurements were performed
separately for charged and neutral  $B$ mesons.

High statistics studies of the hadronic recoil mass $M_X$ in $B \!\to\!
X\,\ell\nu $ provide good means of obtaining an accurate value for
$|V_{ub}|$. The photon spectrum in $B\!\to\! X_s\!+\!\gamma $
decay gives a more or less direct measurement of the heavy quark
distribution function, far beyond indirect constraints on its gross
features inferred from estimates of  a few moments available a few years
ago. The universality relations allow to convert the spectrum into
quantitative estimates of the required fraction $\Phi_{\rm  sl}$ of
$B\!\to\!X_u \,\ell \nu $ decays with high $M_X$ rejected to suppress
the $b\!\to\!c$ background. For example, using the simplest version --
Eqs.~(\ref{90}), (\ref{92}) -- we obtain $\Phi_{\rm  sl}(1.7\GeV) \simeq
0.65$ assuming $\Phi_\gamma(2\GeV)\!=\!12\%$. We have
argued that already the present uncertainty in this $b\!\to\! u$
fraction is not significant, and suggested the way to improve it for
further studies by  incorporating $1/m_b$ effects.

There are statements in the literature that extracting $|V_{ub}|$  from
the $M_X^2$ distribution is seriously plagued by uncertainties in the 
heavy quark distribution function $F(x)$, while decay rates with a cut
$q^2\!>\!(M_B\!-\!M_D)^2$ are accurately calculable and stable  against
higher order effects. Such arguments miss, however, some basics of the 
heavy quark expansion and in particular how the Fermi motion itself
emerges in  the OPE. The claim that some fraction of $\Gamma_{\rm
sl}(b\to u)$ constrained  by a cut on $q^2$ does not depend on $F(x)$ is
actually justified only in a  negative way: this width is governed by a
whole series of unknown expectation  values of higher-order heavy quark
operators which {\it cannot} be resummed or otherwise related  to
$F(x)$. The latter is known reasonably well since it is strongly 
constrained by sum rules and can directly be bounded by data. The
corrections  to widths with a cut in $q^2$ are obscure in this respect;
there is  actually indirect evidence that they are quite significant
\cite{vub}. A  similar reservation was expressed recently in
Ref.~\cite{voloshin}.

To give a brief illustration of missing pieces, let us consider a 
constrained fraction of the $B\!\to\! X_s \!+\!\gamma $ events
\beq
1\!-\!\Phi_\gamma(E) = \frac{1}{\Gamma_{bs\gamma}}\,
\int_E^{\frac{M_B}{2}} {\rm d}E_\gamma \, \frac{{\rm 
d}\Gamma_{bs\gamma}} {{\rm d}E_{\gamma}}\;.
\label{110}
\eeq
Ignoring the perturbative corrections, the usual $1/m_b$ expansion 
always yields the spectrum in the form of expanding around the
free-quark kinematics:
\beq
\frac{1}{\Gamma^0_{bs\gamma}}\,\frac{{\rm d}\Gamma_{bs\gamma}} 
{{\rm d}E_{\gamma}} = a\,\delta(E_\gamma\!-\!\mbox{$\frac{m_b}{2}$}) +
b\,\delta'(E_\gamma\!-\!\mbox{$\frac{m_b}{2}$}) + c\,
\delta''(E_\gamma\!-\!\mbox{$\frac{m_b}{2}$}) + ...
\;,
\label{112}
\eeq
where $a$, $b$, ... are given by the expectation values of local 
$b$-quark operators over $B$. Naively computing $1\!-\!\Phi_\gamma(E)$
in this  way would yield unity for any $E \!>\! \frac{m_b}{2}$ -- the
result clearly unjustified  until `hardness' $M_B\!-\!2E$ is high enough
in hadronic scale. Moreover,  if $\Phi_{\rm sl}(M)$ were computed in the
same naive $1/m_b$ expansion as was used to quantify the corrections to
the $q^2$-constrained semileptonic widths, one would have $\Phi_{\rm 
sl}(M)\!=\!1$ for any $M \!\lsim \!1.8\GeV$, with the decrease only due
to perturbative effects.

Similar reservations apply to the computation \cite{llmw} of 
`incomplete' moments evaluated over the restricted domain $E_\gamma >
2\GeV$  which have been used by CLEO to extract $m_b$. The offered
expressions depend on  the value of the lower cutoff ($2\GeV$) only in
the perturbative piece. Switching off the perturbative corrections,
there would be no dependence on the  cutoff, which cannot be true in
general and thus introduces an additional uncertainty. It is likewise
obvious that the true dependence is  more complicated and involves both
perturbative and nonperturbative  effects in a more intricate way than a
simple sum. This complication would  be reduced if the immediate part of
the spectrum below $2\GeV$ can be incorporated into the analysis.
\vspace*{.1mm}

In radiative $B$ decays one indeed expects only a small fraction of the
spectrum to reside below $2\GeV$, yet not an insignificant amount: we
anticipate about $12\%$. It contributes a barely significant amount to
the overall ${\rm BR}(B \!\to\! X_s \!+\! \gamma)$. Nevertheless its 
accurate measurement is important for various reasons.  We expect
-- and hope -- that new data sets with even higher statistics will
reveal a low energy  spectrum at the predicted level. Detailed study of
this part of the spectrum are  instrumental for a comprehensive program
of exploring local quark-hadron duality; here it  is probed in the
somewhat special case of hard heavy-to-light flavor transitions which
has its own advantages as well as complications. Should it be confirmed,
we expect the (properly defined) heavy quark parameters $\La(1\GeV)$,
$\mu_\pi^2(1\GeV)$ -- and possibly $\rho_D^3$, $\rho_{LS}^3$ -- to be
measured with the comparable precision and to add confidence in our
present knowledge of these parameters. Moreover, combining the whole set
of emerging new data with tight theoretical constraints and precisely
extracted $m_b(1\GeV)$ will allow one to address in a quantitative
fashion higher-order corrections -- a possibility considered so far less
than remote in conventional approaches.

One of the important quantities in this quest is just the integrated
low-$E_\gamma$ fraction of radiative decays $\Phi_\gamma(2\GeV)$ -- even
leaving aside the aspect that it is directly related to the 
determination of $\Phi_{\rm sl}(M)$ at experimentally relevant cuts 
$M \!\approx\! 1.7\GeV$. Finding in the data a fraction of only, say, 
about $5\%$ would tell us that some basic  elements of the description 
had to be revisited. A possible interpretation would be that  processes
involving the emission (or excitation) of ``real'' gluonic degrees of
freedom start to exhibit perturbative local duality at  higher scales
than when only quark degrees of freedom are involved. This actually
represents a conceivable scenario \cite{vadem}. The higher onset of
duality  here would mean that a perturbative treatment of gluon
radiation is not reliable till one reaches a higher scale. From the
specific case under study here with  the separation scale around
$E_\gamma\!\simeq\! 2\GeV$ one would be sensitive to the duality onset
scale near $1.5\GeV$; the lower part of the spectrum would allow to
probe duality for even harder  gluons. The total integrated decay rate
still can be fairly insensitive to such effects. The photon decay
spectrum from this perspective is a  complementary tool to study details
of local parton-hadron duality compared to  hadronic mass distribution
in semileptonic $b\!\to\! c$ decays \cite{vadem}. It  is more ambiguous
as a probe of lower scales due to more significant  interference with
the nonperturbative Fermi motion, yet may improve at lower $E_\gamma$ if
prompt photons and their production mechanism can be reliably
identified.

Although physics of gluon bremsstrahlung is quite different in
semileptonic and radiative $B$ decays, finding a significant  violation
of gluon-hadron duality in the latter at $E_\gamma \!\lsim\! 2\GeV$
would raise concerns in treating, say the hadronic moments in the
former. This especially applies to higher moments in the environment of
present CLEO measurements: the cut on the lepton energy
$E_\ell\!>\!1.5\GeV$  effectively decreases the expansion mass parameter
down from the usual energy  release $m_b\!-\!m_c$.

\vspace*{.2cm}
{\bf Acknowledgements:}~~N.U.\ thanks U.~Aglietti, Yu.~Dokshitzer  and
M.~Misiak for useful clarifications. Discussions with M.~Artuso and
P.~Roudeau are gratefully acknowledged. This work has been  supported in
part by the National Science Foundation under grant number  PHY00-87419.

\end{document}